\documentclass[amsmath,twocolumn,superscriptaddress,prb,aps, reprint]{revtex4}
\usepackage{amsthm,amsfonts,graphicx,verbatim, color}
\usepackage{graphicx}
\usepackage{bm}
\usepackage[utf8]{inputenc}
\let\oldmarginpar\marginpar
\renewcommand\marginpar[1]{\-\oldmarginpar[\raggedleft\footnotesize #1]%
{\raggedright\footnotesize #1}}

\newcommand{\be}{\begin{equation}}
\newcommand{\ee}{\end{equation}}
\newcommand{\bea}{\begin{eqnarray}}
\newcommand{\eea}{\end{eqnarray}}

\newcommand{\Tr}{{\rm Tr}\,}
\renewcommand{\epsilon}{\varepsilon}
\renewcommand{\vec}[1]{{\bf #1}}

\renewcommand{\cite}[1]{[\onlinecite{#1}]}

\def\beq{\begin{equation}}
\def\eeq{\end{equation}}
\def\bea{\begin{eqnarray}}
\def\eea{\end{eqnarray}}

\begin{document}

\title{Spectral features of a many-body-localized system weakly coupled to a bath}
\author{Rahul Nandkishore}
 \affiliation{Princeton Center for Theoretical Science, Princeton University, Princeton, New Jersey 08544, USA}
 \author{Sarang Gopalakrishnan}
 \affiliation{Department of Physics, Harvard University, Cambridge, Massachusetts 02138, USA}
\author{David A. Huse}
 \affiliation{Princeton Center for Theoretical Science, Princeton University, Princeton, New Jersey 08544, USA}
 \affiliation{Department of Physics, Princeton University, Princeton New Jersey 08544, USA}
\begin{abstract}

We study many-body-localized (MBL) systems that are weakly coupled to thermalizing environments, focusing on the spectral functions of local operators.
These spectral functions carry signatures of localization even away from the limit of perfectly isolated systems.
We find that, in the limit of vanishing coupling to a bath,
MBL systems come in two varieties, with either discrete or continuous local spectra.
Both varieties of MBL systems exhibit a ``soft gap'' at zero frequency in the spatially-averaged spectral functions of local operators, which serves as a diagnostic for localization.
We estimate the degree to which coupling to a bath broadens these spectral features, and find that some characteristics of
incipient localization survive as long as the system-bath coupling is much weaker than the characteristic energy scales of the system.
We discuss the crossover to localization that occurs as the coupling to the external bath is tuned to zero.
Since perfect isolation is impossible, we expect the ideas discussed
in this paper to be relevant for 
experiments on many-body localization.
\end{abstract}
\maketitle


\section{Introduction}
Closed quantum many-body systems with quenched randomness can display localization \cite{Anderson}, a phenomenon whereby the system fails to act as its own heat bath and does not approach thermodynamic equilibrium.
The existence of localization in weakly interacting systems has been established to all orders in perturbation theory \cite{agkl, Mirlin,BAA,imbrie}.
Numerical studies \cite{Oganesyan, Znid, pal, Kjall} suggest that such `many-body localization' (MBL) can occur even in strongly interacting systems at high energy
densities, and, indeed, that all the many-body eigenstates of a system can exhibit MBL.
MBL was also shown \cite{LPQO, Bauer, Pekker, Bahri, Vosk, lspt, qhmbl, arcmp} to have many surprising consequences, such as the possibility of symmetry breaking and/or topological order even when such order is forbidden in thermal equilibrium.

Most of the existing literature on MBL assumes the system of interest is \emph{perfectly} isolated from its environment, because the sharp distinction between the MBL and thermal phases only exists in this limit.  In any realistic experiment, however, \emph{some} degree of coupling to an external environment is inevitable. Thus, in order to interpret experiments studying MBL~\cite{basko-expt, ovadyahu, cooper, yao13}, it is imperative to know which features of MBL survive, and in what form, in such imperfectly isolated settings.

In this paper, we note that the spectral functions of local operators retain signatures of many-body localization even in the presence of a
weak system-bath coupling. Such local spectral functions govern transport in an almost MBL system, as well its properties as a quantum memory.
We study the properties of such spectral functions in the regime [called the `fully many-body localized' (FMBL) regime] in which
\emph{every} many-body eigenstate of our isolated system is localized. We first discuss the spectral functions in the limit of vanishing
system-bath coupling, assuming that the thermodynamic limit is taken before the limit of perfect isolation.
We find that FMBL systems then fall
into two categories, depending on whether the
local spectrum at a specific site in a specific sample is continuous or discrete; we term these cases weak and strong MBL respectively. The \emph{spatially averaged} spectra do not show this distinction; however, they universally exhibit a ``soft gap'' at zero frequency that is a diagnostic of localization. We estimate how far these effects are smeared out by a weak system-bath coupling, and argue that manifestations of these effects persist in the spectral functions so long as the system-bath coupling is weaker than the intrinsic energy scales of the system, such that the thermalization timescale is not the shortest timescale in the problem.  We also discuss the crossover to localization that occurs as the coupling to the bath is tuned to zero.

\section{Model}
 It is expected \cite{Lbits, vosk2012, Abanin, Swingle} that the Hamiltonian of an isolated FMBL system can be
written in terms of localized constants of motion (`l-bits' \cite{Lbits}) as follows:
\begin{eqnarray}
H_0  & = &   \sum_ih_iS^z_i + \sum_{i,j}U_{ij}S^z_i S^z_j \nonumber \\
  & & \quad + \sum_n \sum_{i,j,\{k\}}K^{(n)}_{i\{k\} j}S^z_iS^z_{k_1}...S^z_{k_n} S^z_j ~.
 \label{lbit}
\end{eqnarray}
Here, the $\{{\bf S}_i\}$ are the Pauli operators of $N$ localized two-level systems that are `dressed' versions of the local degrees of freedom (`p-bits' \cite{Lbits}; for specificity we assume these p-bits are also two-level systems).
The $\{S^z_i\}$ commute with one another and with $H_0$, so
the eigenstates of $H_0$ are simultaneously also eigenstates of all the $\{S^z_i\}$.  The local fields $h_i$ and the interactions $U_{ij}$, $K^{(n)}_{i\{k\}j}$ are static random variables.  The interactions fall off exponentially with distance, both in their typical values and also in the probability of having a strong interaction.
The $S^z_i$ are related to the p-bits by a system-specific local unitary transformation
whose `kernel' also falls off exponentially with distance \cite{Lbits, Abanin, Bauer}.
We define $\bar h$ to be the typical energy change associated with flipping a single l-bit.

We also define the {\it effective}
two-spin interaction $\tilde U_{ij}$ between two l-bits $i$ and $j$
for each given many-body eigenstate of $H_0$. We define this effective interaction, following \cite{Lbits}, as
\begin{equation}
\tilde U_{ij} = U_{ij} + \sum_{n,\{k\}}K^{(n)}_{i\{k\} j} S^z_{k_1} S^z_{k_2} ... S^z_{k_n}. \label{Ueff}
\end{equation}
We define a decay length $\xi$ for the effective interaction by the exponential decay with distance of the typical magnitude of
$|\tilde U_{ij}|\sim\exp{(-|{\bf r}_i-{\bf r}_j|/\xi)}$.  Note that we do not call this $\xi$ the localization length, since near the many-body localization phase transition it might differ strongly from the length scale of the localization of an l-bit when it is written in terms of the bare p-bits.

The effective interaction between two l-bits a distance $r$ apart in a one-dimensional system is the sum of $\sim 2^r$ terms in the sum in Eq. (\ref{Ueff}).  If these terms are each of random sign, then a typical value of a single term is $\sim 2^{-r/2}\exp{(-r/\xi)}$. In higher dimensions the terms that dominate the sum in (2) will have the l-bits $\{k\}$ all near the straight line segment between sites $i$ and $j$. In any dimension, the decay length of the effective interaction $\xi$ is in general longer than the decay length of an individual
term on the right hand side of (2) \cite{Lbits}. An exception is the limit of weak interactions between localized fermions,
as considered in Ref. \cite{BAA}.  There the two-particle Hartree interaction $U_{ij}$ is the dominant contribution to the effective interaction $\tilde U_{ij}$; both $U_{ij}$ and $\tilde U_{ij}$ fall off with the same $\xi$, which is set by the single-particle localization length and can thus be of any magnitude.
The effective l-bit interaction decay length $\xi$ is the length scale that will be most relevant for the present work.  It should be
possible to measure $\xi$ using the `double electron-electron resonance' technique that is presented in Ref. \cite{DEER}.

 We take the bath to consist of 
 interacting bosons (e.g., anharmonic phonons) hopping on the same lattice as the l-bits \cite{footnoteeth}.
 One possible generic form for the bath Hamiltonian is
\begin{equation}
H_{bath} =  t \sum_{\langle i j   \rangle} b^{\dag}_{i} b_{j} + \Lambda   \sum_{\langle ijk \rangle} (b^{\dag}_{i} b^{\dag}_{j} b_{k} + h.c.) ~.
\end{equation}
We assume that the bandwidth of the bath $t$ is much larger than the characteristic energy scales in the system, so that the bath can
locally supply enough energy for any local process in the system.  In this `broad bandwidth bath' limit,
the energy diffusivity of the bath will be high, such that the bath behaves in an effectively Markovian fashion on the timescales of interest.
The case of a 
narrow-bandwidth bath is discussed in Ref. \cite{meanfield}.
To ensure that this bath remains well-behaved when we consider an infinite temperature, we impose the (artificial)
constraint of no more than some small number (say, two) bosons at any site.
However, we emphasize that our results do not depend qualitatively on the nature of the bath, beyond the assumption that it
thermalizes itself and has a local bandwidth of order $t$. 
Any non-integrable quantum system obeying the eigenstate thermalization hypothesis (ETH) \cite{Deutsch, Srednicki, Rigol} can function as the bath.

The system-bath coupling should be local in the space of p-bits, which implies that it is also local in the space of l-bits.
Here we take the simplest fully local coupling, which has the form
\begin{eqnarray}
H_{int} &=& g \sum_{\langle ij \rangle} 
S^x_i(b^{\dag}_{i } + b_{ i }) ~ \label{Hint}. 
\end{eqnarray}
More generally there should be longer-range and higher-order couplings that fall off exponentially with  distance;
including these would not qualitatively change our results.

For a thermodynamically large bath, any nonzero coupling $g$ should suffice to bring the system to thermal equilibrium,
 in the sense that the new eigenenergies of the coupled system+bath will obey RMT level statistics and the new eigenstates will obey ETH.
More generally, when both the system and bath are finite-sized, the crossover to thermalization occurs when $g\sim \sqrt{t\delta}$,
where $\delta$ is the many-body level spacing of the bath. 
This follows
because each local system-bath coupling term couples to $\sim t/\delta$ other states in the bath, so that the matrix element is typically $\sim g\sqrt{\delta/t}$.
Thermalization will occur when the matrix element becomes of order the many-body level spacing $\delta$ of the bath.

\section{Spectral features in the limit $g\rightarrow 0$}
We now imagine starting from the coupled system and bath, 
and slowly taking the limit $g \rightarrow 0$ so the system remains at equilibrium with the bath.
For specificity, we now consider the spectral function for `flipping' l-bit $j$: 
\begin{eqnarray}
A_j(\omega) & \equiv &\mathrm{Im} \int_{0}^{\infty} dt e^{i \tilde \omega t} \Tr \left[ \rho S^-_j e^{-iHt} S^+_j e^{iHt}\right] \nonumber\\&+&\mathrm{Im} \int_{0}^{\infty} dt e^{i \tilde \omega t} \Tr \left[ \rho S^+_j e^{-iHt} S^-_j e^{iHt}\right],
 \end{eqnarray}
 where $\rho$ is the probability density operator and  
 $H$ is the full Hamiltonian of the system interacting with the bath, $\tilde \omega = \omega + i 0$, the spectral function is measured in thermal equilibrium at a specific site $j$ for a specific disorder realization, and we use units where $\hbar=k_B=1$.  The generalization to spectral functions of other local operators is in principle straightforward.

The qualitative behavior of the spectral function depends on the temperature. At $T=0$ and $g=0$, $A_j(\omega)$ is simply a single delta-function,
corresponding to flipping l-bit $j$ up to its excited state.  At non-zero temperature but still $g=0$, each eigenstate of $H_0$ that has appreciable
Boltzmann weight contributes a delta-function peak, and each such peak is at a different frequency
(as it depends on the states of all the other l-bits); a natural question to ask is whether $A_j(\omega)$, for a specific site $j$ and a specific
realization of the quenched disorder, is discrete or continuous for an infinite system in the limit $g \rightarrow 0$.

\subsection{One dimension}
In one dimension, whether the local spectrum is asymptotically discrete or continuous depends on the decay length $\xi$, which controls how rapidly the effective
interaction $\tilde U_{ij} \sim U_0 \exp(-|i-j|/\xi)$ falls off with distance in a typical eigenstate. One can see this as follows
(assuming for now that $T \rightarrow \infty$):
consider $A_i(\omega)$ for a single l-bit spin $i$ in the middle
of an infinite chain.  When $U_0 = 0$, this contains two delta functions at $\omega = \pm h_i$. Including nearest-neighbor interactions causes these to each split
into four delta functions with typical splittings $\sim U_0 \exp{(-1/\xi)}$.  Including second-neighbor interactions causes each of those delta functions to split on a smaller energy scale
into another four delta functions, etc.  By including these splittings one by one from strongest to weakest, we build up a `spectral tree' (Fig.~1) \cite{Pekker}.  In general, after taking into account interactions with the $2n$ other l-bits within a distance $n$ of site $i$, the original delta
functions have each split into $2^{2n}$ delta functions which are spread over a frequency range that remains finite at large $n$.  Thus the {\it average} gap between
delta functions scales as
$\sim {U_0}  \exp (- n \ln 4)$. 

\begin{figure}[htbp]
\begin{center}
\includegraphics[width =  \columnwidth]{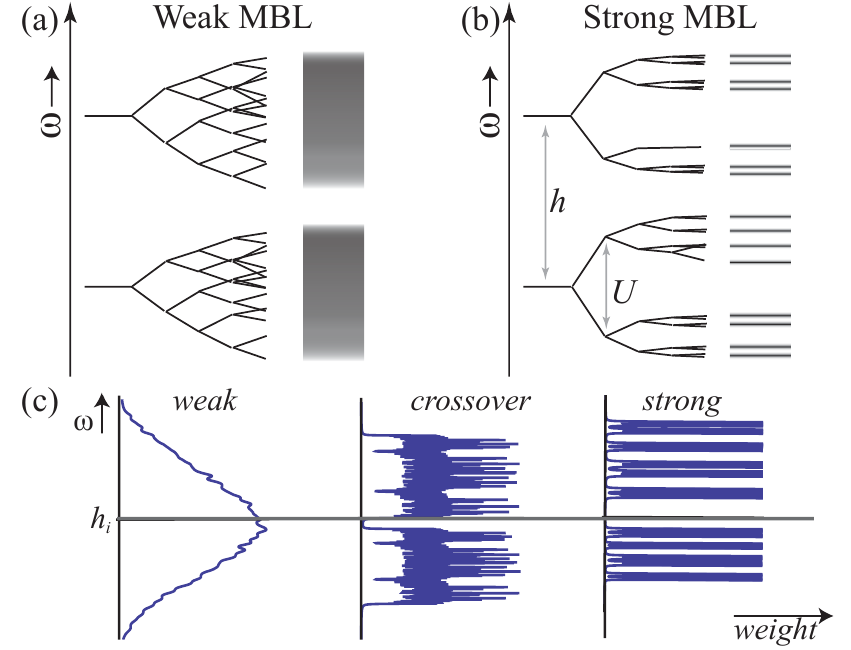}
\caption{Spectral structure at a site $i$ in the weak [panel (a)] and strong [panel (b)] regimes of many-body localization, at infinite temperature. Each configuration of the other spins in the system gives rise to a pair of spectral lines; the resulting spectral structure can be visualized
 as a tree; at the $n$th level of the tree, we have accounted for the influence on site $i$ of all sites out to distance $n$.
 The `true' branching ratio of this tree is more than two; however, we have drawn the tree with a branching ratio of two to avoid clutter.
 (c) A system in the weak MBL regime in one dimension (left) has no sharp features in the spectral function, whereas a system in the strong MBL regime (right) has sharp features at all energy scales. Thus, looking for sharp features at multiple energy scales provides a diagnostic for `strongness' and `weakness,' which survives even away from the limit of perfect isolation. In higher dimensions, the spectral function has sharp features on energy scales greater than $E_c$.}
\label{spectralb}
\end{center}
\end{figure}

Meanwhile, the typical splitting coming from the effective interaction with the spins at a
distance $n$ is $\sim U_0 \exp(-n/\xi)$. For $\xi < 1/\ln 4$ this is much smaller
at large $n$ than the average gap, so opening these gaps causes very few crossings of the `branches' of the spectral tree.
The union of all the new gaps produced due to interactions at large distance $n$ occupies a vanishing fraction of the spectrum.
Thus the gaps opened at any large $n$ do not fill in with spectral weight as $n$ is subsequently increased to infinity, and
the full $n\rightarrow\infty$ spectrum is discrete, with an infinite number of gaps. This can be seen (Fig.~\ref{spectralb}) to lead to an asymptotically pointlike,
statistically self-similar spectrum with a fine structure similar to a Cantor set. We call this the regime of {\it strong} MBL.  For $\xi > 1/\ln 4$, on the other hand,
the new gaps that are opened at any large $n$ generally overlap strongly, causing many crossings of the branches of the spectral tree, and allowing the many delta functions produced for $n\rightarrow\infty$ to densely fill in
almost all gaps.
We call this regime {\it weak} MBL. Note that although local spectra in the weak MBL regime resemble those of a diffusive system in being continuous, they are \emph{not} the same: in particular, as discussed below, they exhibit a soft gap as $\omega \rightarrow 0$.

At any finite temperature $T$, the Boltzmann distribution is dominated by a subset of all eigenstates, with entropy per spin $s(T)<\ln{2}$.  The effective interaction
for a typical thermally-occupied eigenstate will have a decay length $\xi(T)$ that generally depends on $T$.
Thus, after taking into account the interactions with all $2n$ spins within a distance $n$ of site $i$, the original spectral lines split into $\sim \exp(2 n s(T))$
spectral lines.
Comparing this to the typical splitting $U_0 \exp(-n/\xi(T))$ coming from interactions with the spins at distance $n$,
leads us to conclude that for a one-dimensional MBL system 
the weak-to-strong transition occurs at
\begin{equation}
2s(T)\xi(T) = 1~.
\end{equation}
The entropy interpolates between $s(T\rightarrow 0) = 0$ and $s(T \rightarrow \infty) = \ln 2$ per spin, and the decay length is measured in units where the one-dimensional density of spins is unity.  For the local spectral function of an l-bit near the {\it end} of a semi-infinite chain, the weak-to-strong transition is instead at $s(T)\xi(T) = 1$. We note that the distinction between weak and strong MBL is a purely spectral distinction, and is {\it not} associated with any differences in the properties of individual eigenstates.

\subsection{Higher dimensions}
In higher dimensions $d>1$, 
only the `weak MBL' regime can be realized. We show this as follows: use units where the density of of l-bits is unity.
 After interactions with all l-bits within a distance $r$ have been taken into account, the spectrum has $\sim\exp(A r^{d} s(T))$ delta functions, where 
 $A$ is the volume of a $d$-dimensional unit sphere.
 Thus, the average gap scales as  $\sim U_0 \exp(-A r^{d} s(T)) $. At large $r$, this is much smaller than the effective interactions
 $\sim U_0 \exp(-r/\xi(T))$, for any non-zero entropy density $s(T) > 0$. Thus
 only weak MBL can be realized in $d>1$.  For $s(T)\xi(T)\gg 1$ the spectral tree has branch crossings immediately and no gaps
 (except maybe one gap if $\bar h\gg U_0\exp{(-1/\xi)}$).
In the opposite limit $s(T)\xi(T) \ll 1$, the gaps opened due to l-bits at distance $r<r_c$ mostly remain open, while beyond $r_c$, the branches of the spectral tree strongly cross so few additional gaps remain open, with
%
\begin{equation}
r_c \sim \bigg(\frac{1}{\xi(T) s(T)}\bigg)^{1/(d-1)} ~.
\end{equation}
The local spectrum of a  weak MBL system will display a finite hierarchy of gaps, with the smallest gap being of order $E_c \approx U_0 \exp(-r_c/\xi)$.
As $\xi \rightarrow 0, E_c \rightarrow 0$.

The distinction between weak and strong MBL is only sharp in the thermodynamic limit. For a specific finite size system, the local spectrum
is necessarily discrete. The number of gaps that are much wider than the typical many-body level spacing will increase without limit with
system size in the strong MBL regime, but will reach a finite limit in the weak MBL regime.

\subsection{Spatially-averaged spectral functions}
Thus far we have discussed spectral functions evaluated at a single site.  Such strictly local spectral functions could be probed using, e.g., spatially focussed laser spectroscopy.  However, other possible measurements, e.g., a.c. conductivity, probe spatially averaged spectral functions.  Local spectral functions at different sites of this random system will have gaps at different frequencies.  After averaging over spatial position in a thermodynamically large sample, there will in general be no gaps in the (averaged) spectral functions; thus, macroscopic spatial averaging erases the distinction between weak and strong MBL.

However, spatially averaged spectral functions of \emph{local} operators still carry a universal signature of localization,
viz. a soft gap as $\omega \rightarrow 0$.  This soft gap is a consequence of energy-level repulsion in the underlying physical Hamiltonian:
it arises because two many-body eigenstates connected by a local operator (e.g., a spin flip) generically mix. 
Due to this, the probability that an operator which is local in real space produces a transition with energy $\omega$ is suppressed as $\omega \rightarrow 0$,
vanishing as $P(\omega) \sim \omega^\beta$.  There are two regimes of behavior. Deep in the MBL phase, the dominant resonant processes at low $\omega$
 involve only a small number of nearby l-bits, and the exponent $\beta$ is set by the symmetry class from random matrix theory (e.g. $\beta = 1$ for the orthogonal ensemble). Closer to the transition to the thermal phase, rare large ``Griffiths'' regions that are locally conducting give a
smaller $\beta$, which is then a Griffiths exponent \cite{gh,vah}.
In both cases,
there is generically a `soft' spectral gap at zero frequency for any few-particle operator that is local in real space,
in the sense that the spectral weight at low frequencies vanishes as $\omega^{\beta}$ \cite{symmetries}.

While systems with thermal many-body eigenstates also display energy-level repulsion, 
this only occurs on the scale of the many-body level spacing, which
is exponentially small in the system's volume, so the scale of this `soft gap' vanishes in the thermodynamic limit.  
For localized systems, on the other hand, the soft spectral gap width $\tilde \Delta$ remains non-zero even in the thermodynamic limit.
Intuitively, this reflects the fact that physical properties of localized systems are largely insensitive to the addition of distant
degrees of freedom. We emphasize again that the soft gap is a property of {\it spatially averaged} spectral functions.

\section{Finite-$g$ crossovers}
So far, we have considered the $g \rightarrow 0$ limit, in which each spectral function consists of a set of delta-function spikes. When $g > 0$, each spike is broadened into a Lorentzian with width $\Gamma(g)$ (estimated below). The distinction between strong and weak MBL is no longer \emph{sharp}, as gaps on scales $\alt \Gamma(g)$ in the strong MBL phase are smeared out by the line-broadening. However, gaps on scales $> \Gamma(g)$ are filled in only weakly, and thus remain distinguishable. Provided that $\Gamma(g) \ll U_0$, the weak and strong MBL regimes have qualitatively different spectra, with the number of weakly-filled-in gaps increasing sharply as one crosses from one regime to the other (Fig. 1(c)). Similarly, although the spatially averaged spectral weight no longer strictly \emph{vanishes} as $\omega \rightarrow 0$, it is strongly depleted, and should follow a power-law $\sim\omega^{\beta}$ in the frequency range $\Gamma(g) \alt \omega \alt \tilde{\Delta}$.  Thus, local spectra retain signatures of MBL physics even away from the limit of perfect isolation, unlike some other properties of the MBL phase such as the failure of the eigenstate thermalization hypothesis.

\subsection{Estimating the line broadening}
We now estimate $\Gamma(g)$. We begin by considering a {\it non-interacting} localized system, described by an l-bit Hamiltonian that contains only the first term in (1).
In this case, the only processes contributing to the linewidth of l-bit $i$ are those that flip it; from the Golden Rule,
one can estimate the rate of this process as $\Gamma_1(g) \sim g^2/t$.
We now turn to the \emph{many-body localized} case: here, there are additional contributions to the linewidth because the flipping of l-bits
near $i$ causes the effective field $h^{eff.}_i \equiv h_i + \sum\nolimits_j U_{ij} S_j^z + \ldots$ acting on l-bit $i$ to change.
Let us first consider a finite-sized system, which contains $N$ l-bits.  At thermal equilibrium at small $g$, this system has $e^{(s(T) N)}$ thermally-populated many-body eigenstates,
where $s(T)$ is the entropy per l-bit.  We assume the system is large enough so that $s(T) N\gg 1$.
At $g=0$ the local spectral function of l-bit $j$ thus contains $\sim e^{(s(T) N)}$ delta-functions of significant intensity.
At small $g$ each of these many-body
states has a `decay' rate $\Gamma_m(g)\sim N s(T) g^2/t$, since any of the l-bits can flip,
but at low $T$ many l-bits are in their ground state and have a Boltzmann-suppressed probability of flipping up to a high energy state.
Thus the typical spectral line is broadened into a Lorentzian with this width, provided that $g$ and $N$ are small enough so that all of the l-bits in the system interact with each other more strongly than they interact with the bath.

We now proceed to the thermodynamic limit. The interaction between l-bits falls off as $U_0 \exp(-R/\xi)$, where $R$ is the separation between l-bits. For sufficiently large separations $R \agt R_c$, this interaction energy scale becomes smaller than the linewidth,
such that interactions at the scale $R_c$ cannot be resolved within the linewidth $\Gamma_m(g)$. L-bits at distances $>R_c$ should then be treated as part of the bath. 
One can estimate $R_c$ as follows:

\beq
\Gamma_m(g) \approx U_0 \exp(-R_c/\xi) \sim s(T) R_c^d g^2/t ~.
\eeq
This self consistently yields the typical linewidth
\beq
\Gamma_m(g) \sim \frac{g^2}{t} s(T) \xi^d  \ln^d \left(\frac{t U_0 }{g^2 s(T) \xi^d} \right) ~,
\eeq
which parametrically exceeds $\Gamma_1(g)$ at small $g$. The full linewidth is $\Gamma(g) \approx \max\{\Gamma_1(g), \Gamma_m(g)\}$: at low temperatures all the nearby l-bits are thermally frozen, but l-bit $i$ can still decay if excited, and thus the line width saturates to $\sim g^2/t$ in the zero temperature limit.

\subsection{Crossover to localization as $g \rightarrow 0$}
We now discuss the crossover to localization that occurs as we take $g \rightarrow 0$. At $g = 0$, the local spectra
can exhibit a hierarchy of gaps, as we discuss above.
At large $g$ these gaps are all filled in by the line broadening, such that the local spectra exhibit no signatures of MBL. Additionally, the DC conductivity is non-zero at any non-zero $g$. However as we take $g \rightarrow 0$, the line broadening $\Gamma$ vanishes, such that many of the gaps in the $g=0$ spectrum are only weakly filled in. In this regime, the local spectra can exhibit a hierarchy of (weakly filled in) gaps, which is diagnostic of proximity to MBL. This `spectral signature' of MBL persists as long as the line broadening $\Gamma(g)$ is less than the largest energy scale on which the $g=0$ spectrum has gaps. If we examine instead spatially averaged spectra, then the key signature of proximity to MBL is the (weakly filled in) soft gap at zero frequency, which is visible as long as $\Gamma(g)$ is smaller than the width $\tilde \Delta$ of this soft gap.

The preceding discussion has important consequences for the system's lifetime as a quantum memory. The `rounding' of spectral features due to averaging over eigenstates encodes the dephasing rate - the rate at which the in-plane components of the spin $\vec{S}_i$ become uncertain due to interactions between l-bits. This dephasing can in principle be reversed by spin echo procedures. However, coupling to a bath (non-zero $g$) introduces dissipation, which leads to an irrecoverable loss of information. The rate at which `classical' information (encoded in the $S^z_i$) is lost is $g^2/t$.  Meanwhile, the decay rate of the spin echo is given by the line broadening $\Gamma(g)$, which can be parametrically larger than the decay rate of classical information \cite{NMR}.

We note that spin echo measurements may be difficult to perform, particularly in the strongly interacting regime when l-bits have small overlap with the `bare' p-bit operators. It is thus also interesting to consider the system properties as a quantum memory in the absence of spin echo. We assume that even though spin echo is inaccessible, it is possible to write and read information, using some operator $\hat O$ (which may act on multiple l-bits). If the local spectrum of this operator $\hat O$ consists of well separated features with only weak `rounding' due to thermal averaging (or non-zero $g$), then the autocorrelation functions $\langle \hat O(0) \hat O(t) \rangle$ will consist of underdamped oscillations, which decay on a timescale set by the rounding. In this regime, the system can serve as a quantum memory (upto a damping timescale set by the spectral rounding) even if spin echo is unavailable.

\section{From l-bits to p-bits}

Thus far we have restricted ourselves to studying spectral functions of l-bits.  We can extend our conclusions to the bare degrees of freedom (`p-bits') by noting that each p-bit has appreciable overlap with only a small number of l-bits. Thus, the spectral functions of the p-bits are qualitatively the same as those of the l-bits.
In particular, in the limit of small $\xi$, each p-bit overlaps with relatively few l-bits, and therefore the characteristically spiky spectra of e.g. the strong MBL regime should be detectable in real experiments. Moreover, the `soft gap' at zero frequency should be present even in the spectra of p-bit operators, since the origin of this soft gap is level repulsion of p-bit states.
 We note that exact diagonalization on small systems coupled to small baths \cite{mblbath} indicates that the spectral functions of p-bits are indeed sharply different in the localized and delocalized phases, and that the differences persist even at couplings $g$ where the eigenstates are effectively thermal.

We note that studying spectral functions should be instructive even for systems with a many-body mobility edge, when the l-bit construction appears to fail. At energy densities well on the localized side of the many-body mobility edge (which occurs at an extensive energy), we expect the phenomenology of such a MBL system weakly coupled to a heat bath to be analagous to the phenomenology of FMBL systems weakly coupled to a heat bath, which we have discussed in this paper.  The behavior at temperatures near a many-body mobility edge remains an open question.

\section{Conclusions}
We have shown that spectral functions of local operators provide a perspective on many-body localization that remains useful even away from the
(experimentally unrealizable) limit of perfectly isolated systems. In the limit of vanishing system-bath coupling, the behavior of local spectral functions
can be used to categorize MBL states into two kinds, viz. ``weak'' MBL states with continuous spectra and ``strong'' MBL states with discrete local spectra with
a hierarchy of gaps.
Also, we have pointed out that the spatially-averaged
spectral functions generically contain a `soft gap' at zero frequency, which is a universal diagnostic of localization.

Moving away from the limit of perfect isolation, we find that a non-infinitesimal coupling to a bath produces a line broadening $\Gamma(g)$, which erases all structure in the spectral functions on scales less than $\Gamma(g)$. We have calculated the line broadening $\Gamma(g)$, identifying a non-analytic log correction which lies beyond perturbation theory, the consequences of which have been recently explored in \cite{meanfield}. We have discussed the crossover to localization that occurs as we tune $g \rightarrow 0$, and have argued that this behavior should manifest itself not just in spectral functions of l-bits, but also in spectral functions of p-bits.

{\it Acknowledgements:}  We thank Sonika Johri, Ravin Bhatt, John Imbrie, Michael Knap, Ronen Vosk, Ehud Altman and Vadim Oganesyan for useful discussions.
This work was supported in part by the Harvard Quantum Optics Center (SG), and by a PCTS fellowship (RN).  It was initiated at the 2013 Boulder summer school for condensed matter physics.


\begin{thebibliography}{99}
\bibitem[footnote (2014)]{footnoteeth}
We could also consider an integrable bath (e.g. non-interacting bosons).  Many of our results would remain unchanged.  But
real thermal environments are non-integrable, so we prefer to treat that case.
\bibitem[symmetries (2014)]{symmetries}
For systems with other conserved quantities in addition to energy,
we must restrict this to local operators that act within a particular symmetry sector.
Operators that connect different symmetry sectors will not show this soft gap, since there is
no level repulsion between different sectors.
\bibitem{NMR}
In NMR language, the spin flip timescale $t/g^2$ is $T_1$, and the spin echo decay timescale $1/\Gamma(g)$ is $T_2$. The argument presented in IV-A establishes that $T_2$ can be parametrically shorter than $T_1$.
\end{thebibliography}
\end{document}